\documentclass[twocolumn,aps,floatfix,superscriptaddress,pre,showpacs]{revtex4-1}

\usepackage{amsmath}
\usepackage{amsfonts}
\usepackage{amssymb}
\usepackage{graphicx}
\usepackage{color}

\usepackage{graphicx,color}
\usepackage{amsmath,amssymb,amsfonts}
\usepackage{bm}
\usepackage{bibentry, natbib}
\usepackage[T1]{fontenc}
\usepackage[english]{babel}
\usepackage[latin1]{inputenc}
\usepackage{ifthen}
\usepackage{footnote}
\usepackage{subfigure}
\usepackage{setspace}
\usepackage{verbatim}
\usepackage{textcomp}

\definecolor{grey}{rgb}{0.5,0.5,0.5}
\newcommand{\be}{\begin{equation}}
\newcommand{\ee}{\end{equation}}
\newcommand{\rmd}{{\rm d}}
\newcommand{\rme}{{\rm e}}
 
\newcommand{\avg}[1]{\overline{#1}}              
\newcommand{\reff}[1]{(\ref{#1})} 

\newcommand{\change}[1]{\textcolor{black}{#1}}

\begin{document}

\title{Nonmonotonic Effects of Migration in Subdivided Populations}
\author{Pierangelo Lombardo}
\affiliation{SISSA -- International School for Advanced Studies and INFN, via Bonomea 265, 34136 Trieste, Italy}
\author{Andrea Gambassi} 
\affiliation{SISSA -- International School for Advanced Studies and INFN, via Bonomea 265, 34136 Trieste, Italy}
\author{Luca Dall'Asta} 
\affiliation{Department of Applied Science and Technology -- DISAT, Politecnico di Torino, Corso Duca degli Abruzzi 24, 10129 Torino, Italy}
\affiliation{Collegio Carlo Alberto, Via Real Collegio 30, 10024 Moncalieri, Italy}

\date{\today}

\begin{abstract}
The influence of migration on the stochastic dynamics of subdivided populations is 
still an open issue in various evolutionary models.
We develop here a self-consistent mean-field-like method 
in order to determine the effects of migration on relevant
nonequilibrium properties, such as the mean fixation time.
If evolution strongly favors coexistence of species (e.g., \emph{balancing selection}),
the mean fixation time develops an unexpected minimum as a function of the migration rate.
Our analysis hinges only on the presence of a separation of time scales between local and global dynamics and therefore it carries over to other nonequilibrium processes in physics, biology, ecology, and social sciences.
\end{abstract}


\maketitle

Natural populations are often subdivided and fragmented in space, 
with the consequence that species or genetic traits 
get locally extinct and recolonized by migration. 
Understanding and predicting how migration among subpopulations affects their collective evolution is therefore an important issue 
across various disciplines, e.g., conservation ecology \cite{hanski}, population genetics \cite{hamilton}, evolutionary game theory \cite{nowak}, language competition \cite{blythe}, learning dynamics \cite{realpe}, and epidemics \cite{colizza}. 

The dynamics of subpopulations results from the competition between 
the evolutionary ``force'' (\emph{selection}) which favors stronger genotypes and the intrinsic noise (\emph{genetic drift}) due to death and reproduction of individuals.
This noise eventually drives any finite population into an absorbing state (\emph{fixation}), in which all individuals have the same traits (e.g., species/language/opinion). 
In subdivided populations, migration acts with selection and internal noise, influencing the statistical properties of the fixation process, such as the {\em mean fixation time} (MFT). In this respect, it is widely accepted that in the absence of spatial embedding,
the effect of subdivision in populations of constant and equal size effectively amounts at a
rescaling of the relevant parameters of the population, such as the population size and the
effective strength
 of the selection \cite{maruyama,slatkin}. 
When selection is constant or absent, the MFT monotonically decreases upon increasing the migration rate \cite{whitlock,cherry-wakeley,blythe}, but more complex behaviors cannot be ruled out a priori.
Here we consider evolutionary forces that favor biodiversity, i.e., the coexistence of species or different genotypes, showing that the MFT can, in fact, display a nonmonotonic dependence on the migration rate. 
Even in the absence of mutation, 
this kind of evolutionary forces
are common in the evolution of natural populations. For instance, 
the so-called
balancing selection
\cite{hamilton,balancing} 
%--- also known as over-dominance %or heterozygote advantage 
%--- 
acts in several contexts, most  
notably mammalian \cite{hcc} and plants \cite{sil}. The maintenance of some genetic diseases in humans, e.g., 
sickle-cell anemia \cite{anemia}, thalassemia \cite{thalassemia} and cystic fibrosis \cite{fibrosis} is also ascribed to balancing selection. Analogous mechanisms are responsible for 
cooperative
behaviors in ecology and coevolutionary dynamics \cite{egt,traulsen}, such as those recently observed in microbial communities \cite{xavier}, or for
% the emergence of 
emergent bilingualism %from 
in language competition \cite{abrams}.

For concreteness, we focus here on a model specific to population genetics, and we investigate the effect on MFT of the interplay between balancing selection and subdivision. We develop a self-consistent mean-field-like approach which yields 
an effective dynamic equation, from which we derive the nonequilibrium collective properties, such as the MFT. For weak selection, our approximation renders the one of Ref.~\cite{cherry-wakeley}.  
We show that the MFT can actually develop a minimum as a function of the migration rate for sufficiently strong selection. This is in contrast to the assumptions in Ref.~\cite{slatkin} and to the intuitive idea that the collective fluctuation needed to reach global fixation could be facilitated by increasing the migration.
The existence of this minimum depends, inter alia, on the optimal frequency, i.e., on the amount of biodiversity promoted by balancing selection alone. 
The nonmonotonicity of the MFT is reflected in the behavior of the 
so-called
 ``heterozygosity'', which quantifies the biodiversity within the subdivided population.

\emph{The model. ---} 
Inspired by common models in population genetics, we consider $\Omega\gg 1$ 
individuals carrying a single copy of a gene with two possible values (\emph{alleles}) $A$ and $B$.
The evolution of  this large but finite population
turns out to
 be effectively described by a \emph{diffusion approximation} \cite{kimura,nota}, i.e., by a Langevin equation for the frequency $x$ of, e.g., allele $A$. 
The mean change of $x$ in a 
well-mixed
 population is $\mu(x) = \tilde{s} x (1-x)$, where $\tilde s$ is the selection rate, while the variance is approximately given by $v(x) = x(1-x)/(\Omega\tau_g)$, where $\tau_g$ is the generation time (see Ref.~\cite{sm} for a derivation of these expressions from microscopic models).
Hereafter, time is measured in units of generations, so that $\tau_g=1$ and the rates become dimensionless quantities.
Balancing selection is characterized by $\tilde{s} = s(x_{\ast} - x)$, where $s>0$ is a constant and $x_{\ast}$ represents the internal optimal frequency which is promoted by balancing effects in an infinite population.

\begin{figure}[t] 
\includegraphics[width=0.95\columnwidth]{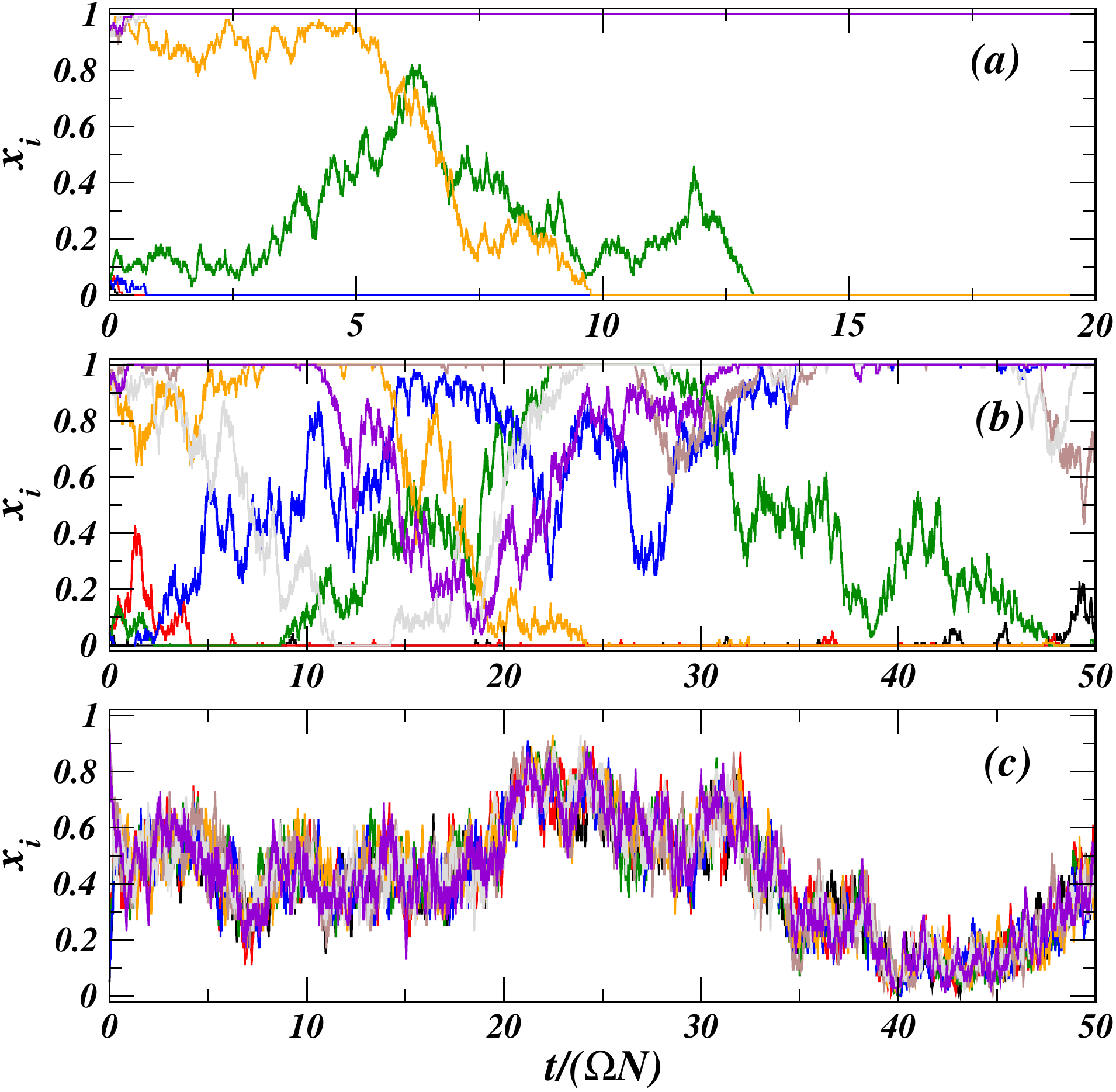}
\caption{Time evolution of the frequency $x_i$ of allele $A$ in the various demes 
\change{(represented by different colors)}
of a fully-connected population consisting of $N=8$ demes with $\Omega=100$ individuals each (a) in the absence of migration ($m=0$) or (b) for small $(\Omega m=0.05)$ and (c) large $(\Omega m=50)$ migration rate. The balancing selection is characterized here by $x_\ast=0.5$ and $\Omega s=5$. At time $t=0$, half of the demes have $x_i=0.05$, while the remaining ones $x_i=0.95$. 
Upon increasing $N$, the fluctuations of $\bar x$ around $x_*$ reduce significantly in panel (c).}
\label{fig:evolution}
\end{figure}

In order to investigate the influence of migration on subdivided populations with balancing selection in the simplest possible setting, we consider the celebrated Island model, originally proposed by Wright \cite{wright} for neutral evolution. It consists of $N$ subpopulations (\emph{demes}), each 
composed by $\Omega$ individuals which evolve as described above (with the same $\mu(x)$ and $v(x)$),
while being allowed to \emph{exchange} a randomly picked individual with any other deme at a rate $m/N$, such that 
$\Omega$ is unchanged.
For sufficiently large $\Omega$ and small $m$ and $s$, the evolution of the allele frequency $x_i\in [0,1]$ in the $i$-th deme is described
by the Langevin equation \cite{kimura,sm} (with It\^{o} prescription), 
\be
\label{Langevin_single-deme}
\dot{x}_i =\mu(x_i)+ m (\bar x - x_i) +\sqrt{v(x_i)}\ \eta_i,
\ee
where $\eta_i$ are independent Gaussian noises with $\langle\eta_i(t)\eta_j(t')\rangle=\delta_{i,j}\delta(t-t')$; 
hereafter the overbar denotes interdeme averages, e.g., $\avg{x^k} =\sum_i x_i^k/N$, and thus $\bar x$ is the interdeme mean frequency (IDMF).
For $m=0$, the demes are independent: the deterministic selection term $\mu$ in Eq.~\eqref{Langevin_single-deme} drives $x_i$ towards $x_{\ast}$, while the random genetic drift 
%eventually
finally
drives $x_i$ towards one of the two possible absorbing states $x_i=0$ and $1$, corresponding to fixation of allele $B$ and $A$, respectively (see Fig.~\ref{fig:evolution}(a)). 
For $m>0$, migration acts as a source of biodiversity for the subpopulations, preventing their independent fixation  
(see Figs.~\ref{fig:evolution}(b) and \ref{fig:evolution}(c)) and favoring a coordinate evolution of the interacting demes. 
For $\Omega m \gg 1$ and $x_*$ sufficiently close to $0$ or $1$, the collective evolution rapidly drives all demes into the same absorbing state; 
instead, for a wide range of parameters, the IDMF $\bar x$ fluctuates for a long time around a value $\hat x$ --- characterized by the vanishing of the deterministic force in the  dynamics of $\bar x$ --- until fixation eventually occurs through a rare (for large $N$) fluctuation 
\change{\cite[Sec.~IIB]{sm}.}
This coordinated behavior around $\hat x$ becomes effectively a {\em metastable} state if the typical time $T_{\rm rel}$ required to reach it from the initial condition is significantly shorter than the typical time $T_{\rm fluct}$ for fixation to occur. This condition is satisfied for $m s \Omega^2N\gg1$ 
\change{\cite[Sec.~IIA]{sm}.}
The statistics of fixation can be studied by considering the evolution equation of $\bar x$, which follows from Eq.~\reff{Langevin_single-deme}, 
\begin{equation}
\label{Langevin_xbarra}
\dot{\avg{x}} =s[x_{\ast} \avg{x}-(1+x_{\ast})\avg{x^2} +\avg{x^3}] +
\sqrt{(\avg{x}-\avg{x^2})/(\Omega N)}\, \eta,
\end{equation}
where $\eta$ is a Gaussian noise with $\langle\eta(t)\eta(t')\rangle = \delta(t-t')$. This equation involves higher-order moments, and the hierarchy does not close; however, we can proceed by introducing a moment closure scheme based on a self-consistent mean-field-like approximation. 
%
%%%
%%%
%

\emph{The approximation. ---}
Since the global variable $\bar x$ is the average of $N$ local frequencies, it is heuristically expected that its dynamics is much slower than that of the individual $\{x_i\}$, determining a separation of time scales between the local and global dynamics. In the absence of selection ($s=0$), Eq.~\eqref{Langevin_xbarra} is driven only by the genetic drift, therefore the time scale separation occurs for sufficiently large $N$ ($N\Omega m\gg1$).
Being coupled only via the slowly varying quantity $\bar x$, $\{x_i\}$ can be considered as almost independent random variables, each one described by a conditional quasi-stationary distribution $P_{\rm qs}(x_i|\bar x)$. The latter can be obtained by solving the stationary Fokker-Planck equation associated with Eq.~\eqref{Langevin_single-deme}, in which $\bar{x}$ is treated as a constant parameter.
Under these assumptions the population average $\avg{x^k}(t)$ can be approximated, for $N\gg 1$, by the corresponding mean $\int \rmd x_i\,x_i^k\,P_{\rm qs}(x_i|\bar x)$.
For $s=0$ one obtains $P_{\rm qs}(x| \bar{x}) \propto  x^{2m'\bar{x}-1}(1-x)^{2m'(1-\bar{x})-1}$, where $m'=\Omega m$ is a rescaled rate introduced for convenience and  $P_{\rm qs}(x| \bar{x})$ satisfies the consistency condition $\bar{x} = \int_0^1 \rmd x\,x\,P_{\rm qs}(x|\bar{x})$. 
This %distribution 
$P_{\rm qs}(x|\bar x)$ can then be used for evaluating $\avg{x^2}$ and  $\avg{x^3}$
%the higher-order moments contained 
in Eq.~\eqref{Langevin_xbarra} and for calculating the mean drift $M(\bar{x})$ and variance $V(\bar{x})$ of the (stochastic) variable $\bar{x}$ \cite{cherry-wakeley}:
\be
\label{cherry}
 M(\bar{x})=s_{\rm e}\,\bar x(1-\bar x)(\change{x_{\ast}^{\rm e}}-\bar x)\ \mbox{and}\ 
 V(\bar{x})=\frac{\bar x(1-\bar x)}{N_{\rm e}}.
\ee
This implies that at the lowest, non-vanishing order in $s$, the subdivided population behaves like a well-mixed one with an \emph{effective selection coefficient}  $s_{\rm e}=s/\left[\left(1+\frac{1}{m'}\right)\left(1+\frac{1}{2m'}\right)\right]$, an \emph{effective population size} $N_{\rm e}=N\Omega\left(1+\frac{1}{2m'}\right)$,
\change{and an \emph{effective optimal frequency} $x_*^{\rm e}=x_*+(x_*-1/2)/m'$}.
The time scale $T_{\rm migr}$ associated with the response of $x_i$ to a variation of $\bar x$ can be read from Eq.~\reff{Langevin_single-deme} and it is $T_{\rm migr}\simeq 1/m$. The typical time scale of the dynamics of $\bar x$ is determined, instead, either by the time scale $T_{\rm rel}\simeq 1/s_{\rm e}$ of the drift or by the time scale $T_{\rm fluct}\simeq N_{\rm e}$ of the stochastic term in Eq.~\reff{cherry}. When $T_{\rm rel} <T_{\rm fluct}$, i.e., $N\Omega s>1+1/m'$, our approximation requires $T_{\rm rel}\gg T_{\rm migr}$, i.e., $s_{\rm e}\ll m$, while in the opposite case, it is accurate whenever $N\gg 1$ (see 
\change{Ref.~\cite[Sec.~IIA]{sm}}
 for a detailed discussion).
This approximation can be generalized to small but non-vanishing values of $s_e/m$ by accounting (a) for $s\neq 0$ in the quasi-stationary distribution $P_{\rm qs}$
and (b) for the fact that $\bar x$ slowly changes during the fast evolution of $x_i$, which results in a distribution $P_{\rm qs}(x_i|y(t))$ where the \emph{effective field} $y(t)\simeq\bar x(t)$ has to be determined self-consistently.
The single-deme quasi-stationary distribution for $s'\equiv\Omega s \neq 0$ is
\be\label{Peq}
 P_{\rm qs}(x|y) \propto  x^{2m' y-1}(1-x)^{2m'(1-y)-1}\rme^{s' x(2x_{\ast}-x)}.
\ee
The consistency condition $\bar{x} = \int_0^1 \rmd x\,x\,P_{\rm qs}(x|y)$ gives $y = \bar{x} - (s_{\rm e}/m) \bar{x}(1-\bar{x})(\change{x_\ast^{\rm e}} - \bar{x}) +O((s_{\rm e}/m)^2)$, which can be used together with Eq.~\reff{Peq} in order to calculate higher-order corrections  in $s$  to $M(\bar{x})$ and $V(\bar{x})$ 
\change{\cite[Sec.~III]{sm}.}
% 
%
%%%
\begin{figure}[t] 
\includegraphics[width=0.92\columnwidth]{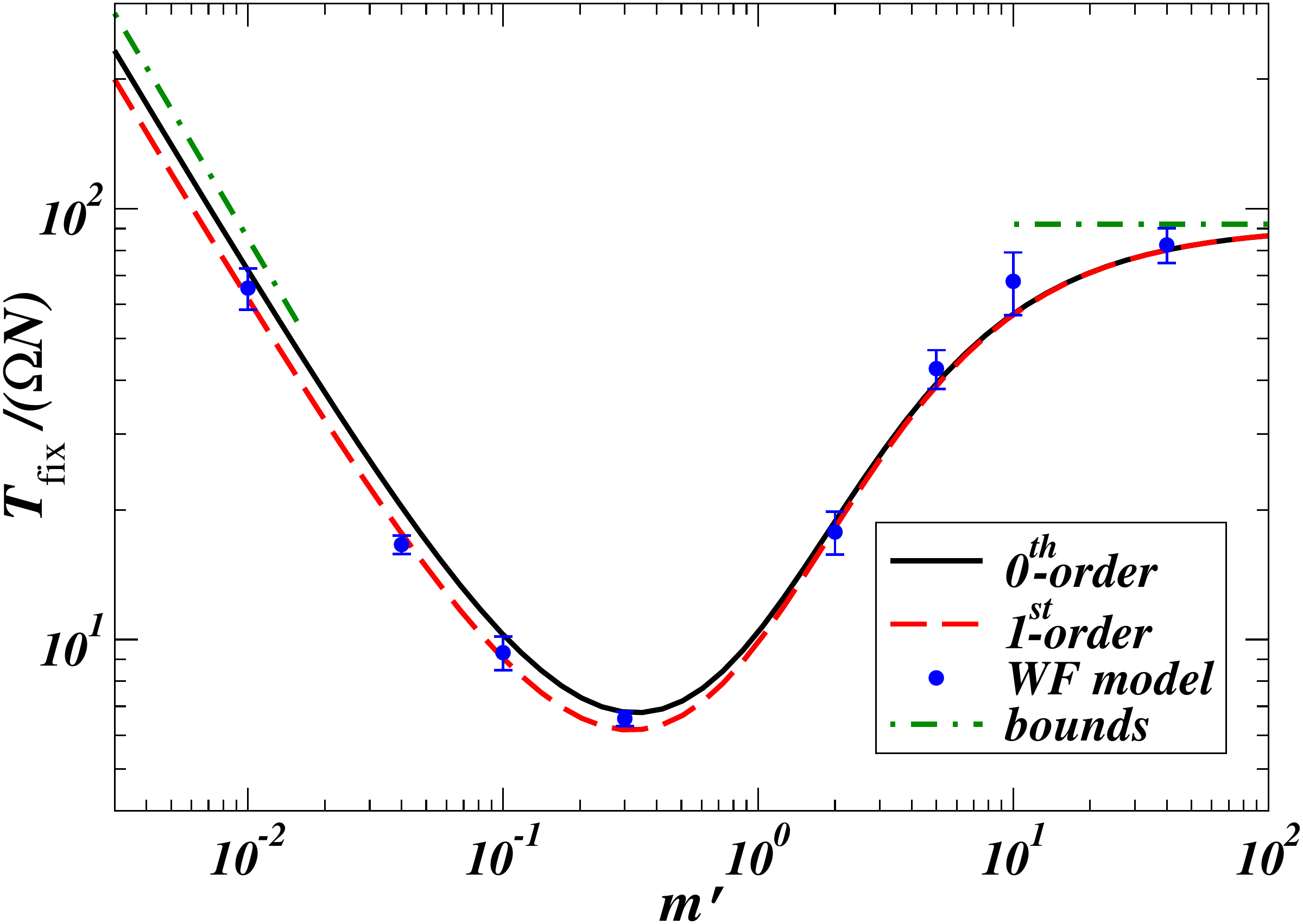}
\caption{Mean fixation time as a function of the migration rate $m'$ 
with $N=30$, $\Omega=100$, $s' = 1$, and $x_\ast=0.5$. 
The solid line corresponds to Eq.~\reff{Tfix0}, while the dashed line accounts for the first-order correction in $s_{\rm e}/m$; symbols with errorbars are the results of numerical simulations of the Wright-Fisher (WF) model. %with selection. %zip
The dash-dotted lines indicate the upper bounds for small and large migration, found in Ref.~\cite{slatkin}. %(see also Ref.~\cite{sm}).  %zip
}\label{fig:Tfix}
\end{figure} 
%%%
%

\emph{Mean fixation time. ---} On the basis of $M(\bar{x})$ and $V(\bar{x})$ calculated as discussed above, the MFT $T_{\rm fix}(\avg{x})$ for the whole population with an initial IDMF $\bar x$ is determined within the diffusion approximation by $V(\bar{x}) T_{\rm fix}''(\bar{x})/2 + M(\bar{x})T_{\rm fix}'(\bar{x}) = - 1$  \cite{kimura-ohta}.
For $x_*=1/2$, by using the lowest-order approximations [${}^{(0)}$]
for $M$ and $V$ in Eq.~\eqref{cherry} and choosing the state $\bar x=1/2$ (corresponding to the metastable state) as initial condition, we get
\be\label{Tfix0}
 T_{\rm fix}^{(0)}=N_{\rme}\int_0^1 \rmd y \int_0^1 \rmd z \frac{\rme^{s_{\rme}N_{\rme} y (1-z^2)/4}}{1- y z^2},
\ee
which reaches a constant value for $m'\gg 1$, while $T_{\rm fix}^{(0)}/(N\Omega) \simeq \log 2/m'$ for $m' \ll 1$.
Figure~\ref{fig:Tfix} shows $T_{\rm fix}^{(0)}$ (solid line) as a function of $m'$  for the population specified in the caption,  together with the prediction (dashed line) 
which accounts for 
%obtained by including 
the first-order correction in $s_{\rme}/m$ to the mean drift $M(\bar{x})$ and variance $V(\bar{x})$ \cite{sm}.
$T_{\rm fix}^{(0)}$
%The MFT 
shows a marked nonmonotonic dependence on the migration rate $m'$, while  complying with the bounds of Ref.~\cite{slatkin} for small and large $m'$ (dash-dotted lines).  
In fact, $T_{\rm fix}^{(0)}(m'\gg 1)$ approaches the value it would have in a well-mixed population of $\Omega N$ individuals, whereas for $m' \ll 1$  fixation --- and thus $T_{\rm fix}$  --- is controlled by the growing time scale $T_{\rm migr}\propto 1/m'$  associated with migration.
In this respect, the limit $m'\to 0$ differs essentially from the case  $m'=0$, in which $T_{\rm fix}$ is governed by the single-deme fixation times, is finite, and it scales $\propto\log N$ for large $N$
\change{~\cite[Sec.~IV]{sm}.}

In order to demonstrate the accuracy of our analytical predictions, Fig.~\ref{fig:Tfix} reports the results (symbols with errorbars)  of numerical simulations of the Wright-Fisher (WF) microscopic model with balancing selection \cite{sm}.
 Their agreement with the analytical prediction of Eq.~\reff{Tfix0} is very good and further improves upon including the first-order corrections in $s_{\rm e}/m$  (dashed line).

\begin{figure}[t]
\includegraphics[width=0.95\columnwidth]{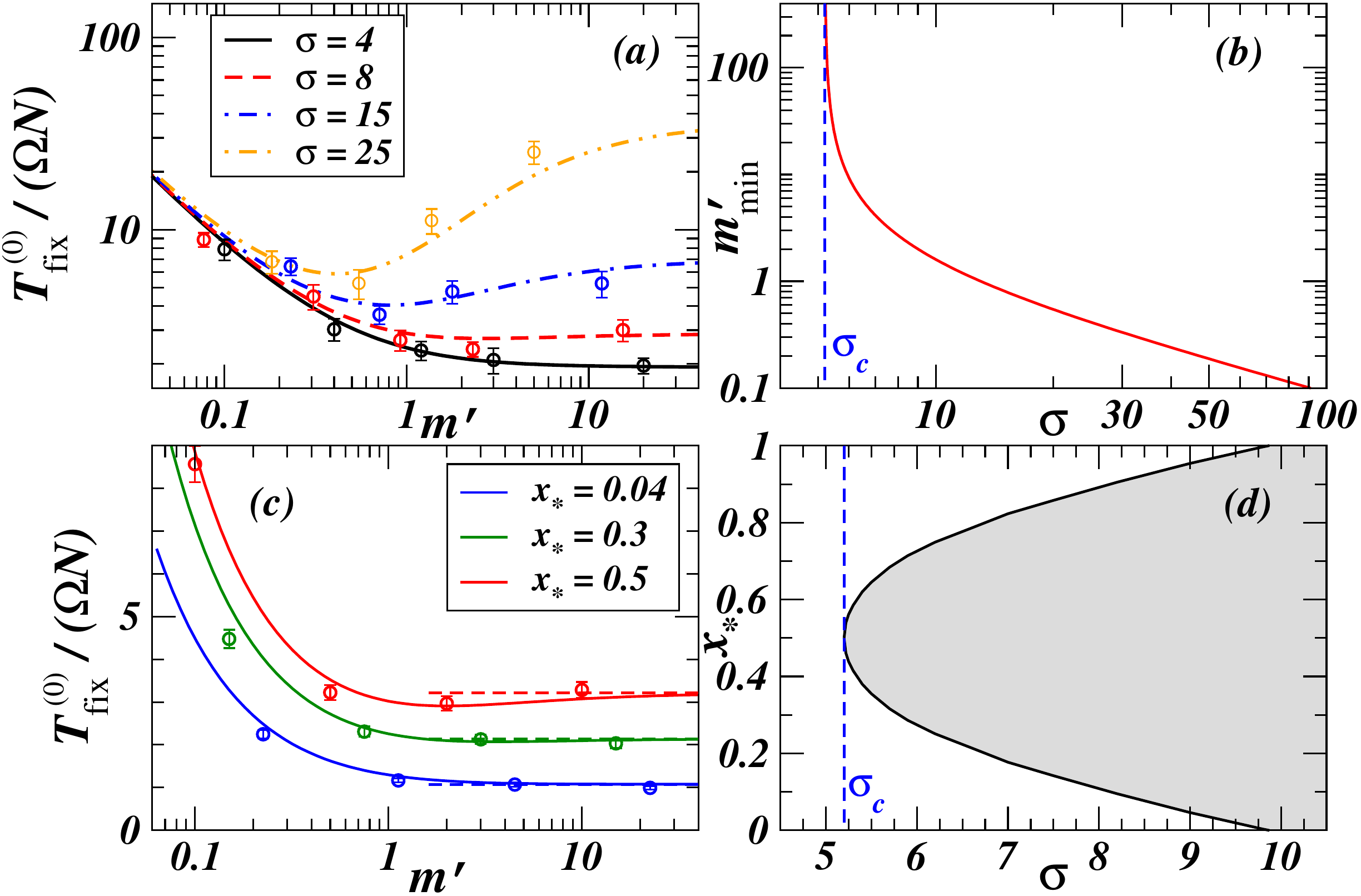}
   \caption{Features of the mean fixation time $T_{\rm fix}^{(0)}$ in Eq.~\reff{Tfix0} for a population of $N=30$ demes with $\Omega=100$ individuals each: (a) dependence of $T_{\rm fix}^{{(0)}}$ on  $m'$ for $x_\ast=0.5$ and various values of $\sigma$; (b) $m'_{\rm min}$ as a function of $\sigma$; (c) $T_{\rm fix}^{{(0)}}$ as a function of $m'$ for \change{$\sigma = 9$} and various $x_\ast$; (d) region \change{(gray)} of the parameter space $(\sigma,x_\ast)$ 
   where %within which
   $T_{\rm fix}^{{(0)}}$ is a nonmonotonic function of $m'$.
   \change{Symbols with errorbars are the results of numerical simulations of the WF model.}
   }   \label{fig:Tfix2}
\end{figure}

Figure \ref{fig:Tfix2}(a) shows that the 
nonmonotonicity displayed in Fig.~\ref{fig:Tfix} is enhanced upon increasing $\sigma \equiv s'N$, while it disappears for $\sigma<\sigma_c$, where $\sigma_c$ is a critical threshold below which the MFT behaves qualitatively as in a neutral population with $s=0$.  
The value $m'_{\rm min}$ of $m'$ at which $T_{\rm fix}$ is minimum diverges for $\sigma \to \sigma_c\simeq 5.2$ and decreases upon increasing $\sigma>\sigma_c$, as shown in Fig.~\ref{fig:Tfix2}(b). The value $\sigma_c$ slightly depends on $s_{\rm e}/m$ if the corrections to Eq.~\reff{cherry} are included. Figure \ref{fig:Tfix2}(c) shows that  the nonmonotonicity of $T_{\rm fix}^{(0)}$ persists also for $x_{\ast}\neq 1/2$, but only within an interval of values of $x_\ast$ which depends on $\sigma$ --- as indicated by the shaded area in Fig.~\ref{fig:Tfix2}(d) --- and which 
\change{covers the entire range for $\sigma\gtrsim 10$.}
%
%%%
%%%
%

\emph{Biodiversity. ---}
Migration is expected to affect the level of biodiversity of  a population. 
In diallelic models, this effect is usually studied in terms of (i) the  \emph{global heterozygosity} $H = 2\bar x(1-\bar x)$, which quantifies the diversification of the global population but neglects the possible subdivision in demes, and of (ii) the  \emph{intra-deme heterozygosity} $h=(2/N)\sum_{i=1}^Nx_i(1-x_i) = 2\avg{x(1-x)}$, which measures the average level of diversification inside each deme.
Note that $0 \leq h \leq H\leq 1/2$. $H=0$ corresponds to the loss of global biodiversity, namely all individuals within the population have the same genotype; $H=1/2$, instead, corresponds to the maximal possible global biodiversity in which the two genotypes are equally present within the whole population.
Analogous interpretation holds for $h=0$ and $h=1/2$ at the intra-deme level.
As depicted in Fig.~\ref{fig:evolution}(c) the local allele frequencies $\{x_i\}$ approach each other for $m'\gg 1$, with $x_i\simeq x_j$ and therefore $h\simeq H$. In the case of moderate migration rate $m'\lesssim 1$ of Fig.~\ref{fig:evolution}(b), instead, different demes fix different alleles, causing $h\simeq 0$, while $H$ is maintained positive by migration which acts as a constant source of biodiversity.

\begin{figure}[t]
\includegraphics[width=0.95\columnwidth]{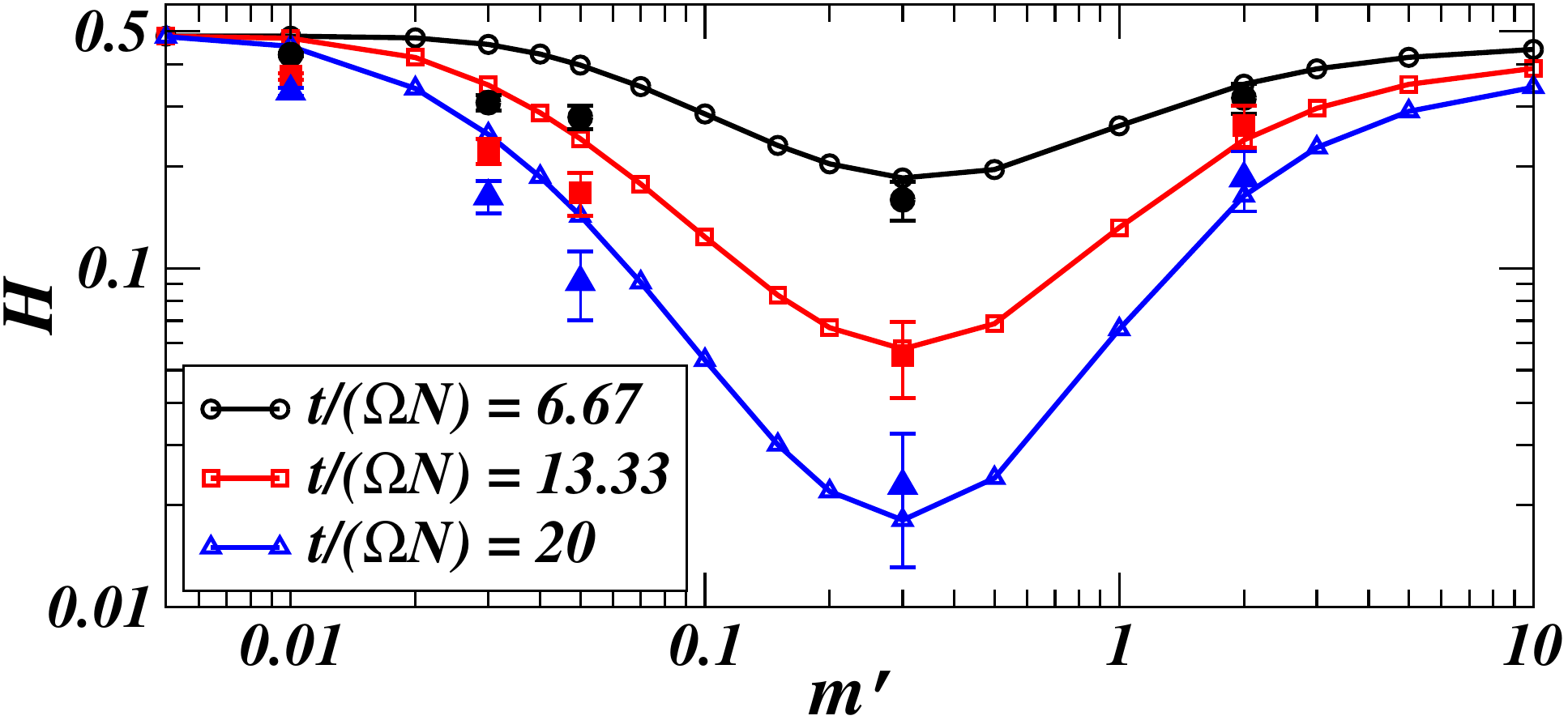}
\caption{Dependence of the global heterozygosity $H$ on the migration rate $m'$ at various times, for a subdivided population with $\Omega=100$, $N=30$, $s'=1$, and $x_* = 1/2$: the prediction of the approximation described in the text (solid lines) is compared with the results of simulations of the WF model (symbols with error-bars).}
\label{h(x0)}
\end{figure}

In order to understand how migration affects biodiversity before the eventual fixation $H=h=0$, we assume that the population at time $t=0$ is in the metastable state $\bar{x} =\hat x$ such that $H(0)=2\hat x(1-\hat x)$ and that it persists in this state until fixation occurs. Under this heuristic assumption, one can approximate $H(t)\simeq [1-p_{\rm fix}(\hat x,t)]H(0)$, where $p_{\rm fix}(x_0,t)$ is the probability that a population prepared with $\bar x=x_0$ at time $t=0$ has already fixed at time $t$.
$p_{\rm fix}$ 
satisfies the backward Fokker-Planck equation $\partial_t p_{\rm fix} = M(x_0)\partial_{x_0} p_{\rm fix} + V(x_0) \partial_{x_0}^2 p_{\rm fix}/2$, which can be integrated numerically.
By using the expressions of $M$ and $V$ in Eq.~\eqref{cherry}, the results of this approximation for $H$ are presented in Fig.~\ref{h(x0)} as functions of $m'$ for some values of 
$t$ and they are compared with those of numerical simulations of the WF model (symbols with errorbars) \cite{sm}.
Note that the estimate of $H(t)$ is expected to become less accurate as  $m'\sigma$ exceeds 1 because, correspondingly, the state $\bar x\simeq \hat x$ is no longer metastable 
\change{\cite[Sec.~IIB]{sm}.}
 For slow and fast migration $H(t)\simeq H(0)$ for a rather long time whereas $H(t)$ rapidly decreases in time for intermediate values of the migration rate.
For a fixed time and as a function of $m'$, instead, $H$ has a minimum at $m'\simeq m'_{\rm min}$, indicating that the global biodiversity can be enhanced upon increasing migration \cite{many-alleles}.
Our predictions agree rather well with the results of simulations, apart, as expected, from $m'\lesssim 1/\sigma\simeq 0.03$. 
A similar study of both $H$ for different values of the parameters and $h$ \cite{sm}
highlights a nonmonotonic dependence on $m'$ whenever the corresponding  $T_{\rm fix}$ develops a minimum.
%
%%%
%%%
%

\emph{Conclusions. ---}
Focusing on the Island model \cite{wright}, we have shown that the mean fixation time of a subdivided population can become a nonmonotonic function of the migration rate $m$ in the presence of balancing selection, an evolutionary mechanism which promotes the coexistence of different genetic traits within the same populations. The emergence of a minimum depends on both the selection strength $\sigma \equiv s\Omega N$  exceeding a threshold and on the frequency $x_\ast$ of coexistence which is promoted by the selection. 
While the MFT increases upon decreasing $m$ because of the slowing down in the migration dynamics, its possible increase for sufficiently large $m$ has a less intuitive explanation.
\emph{A posteriori} this is due to the formation of a metastable state, the ``life time'' of which might increase upon increasing the migration rate.
Our result extends beyond population genetics: 
it carries over to any other evolutionary model whose dynamics has an internal attractive equilibrium (coexistence) in addition to absorbing states (specialized states).
Moreover, these features should also appear in subdivided populations with more complex migration or spatial  \cite{korolev} structures.
It would be interesting to understand whether the features discussed here also emerge by introducing balancing selection in those population models
for which subdivision induces a bifurcation \cite{altrock}, a phase transition \cite{waclaw}, or a maximum in some characteristic times of the dynamics \cite{khasin}.
The approach presented here for describing the dynamics of the entire population via an effective Langevin equation 
can be generically applied to any collective dynamics in which fast local variables are influenced by slow, global, ``mean-field-like" quantities.
In this respect, it extends to transient properties the self-consistent mean-field-like approximations typically used in statistical physics to investigate the stationary properties of nonequilibrium processes \cite{vandenbroeck}.

\emph{Acknwoledgments. }L.D.~acknowledges the Italian FIRB Project No. RBFR10QUW4.

% voglio evitare che la bibliografia sia indicata nella table of contents
\begingroup
\renewcommand{\addcontentsline}[3]{}% Remove functionality of \addcontentsline
\renewcommand{\section}[2]{}% Remove functionality of \section

\endgroup

%%%%%%%%%%
%%%%%%%%%%
%%%%%%%%%%
%%%%%%%%%%
%%%%%%%%%%
%%%%%%%%%%
%%%%%%%%%%
%%%%%%%%%%
%%%%%%%%%%
%%%%%%%%%%

\newpage

\begin{widetext}

% voglio evitare che il titolo Supplemental Material nella table of contents
\begingroup
\renewcommand{\addcontentsline}[3]{}% Remove functionality of \addcontentsline
\renewcommand{\section}[2]{}% Remove functionality of \section

\subsection*{\huge Supplemental Material}

\endgroup

\tableofcontents

\section{Derivation of the single-deme Langevin equation}
 
Here we show that
the microscopic Wright-Fisher and Moran models are accurately described, for
large populations and small selection and migration rates, by the Langevin equation which is discussed in the Letter.

\subsection{Wright-Fisher model}\label{WFsection}

The Wright-Fisher model \cite{sFisher,sWright}, consists of a (haploid) population of $\Omega$ individuals, each of which carries one of two possible alleles $A$ and $B$. At each time step of the dynamics --- corresponding to reproduction ---  the whole population is replaced by a new generation in which the allele of each new individual is drawn at random with a probability determined by the properties of the previous generation. The time interval $\tau_g$ between two consecutive steps represents the duration of a generation and hereafter we set $\tau_g=1$. In a neutral model, i.e., in the absence of selection, each new individual is chosen to carry allele $A$ (resp. $B$) with probability  $x=\Omega_A/\Omega$ (resp. $1-x$), where $\Omega_A$ indicates the number of individuals carrying allele $A$ in the previous generation.
A difference in allele fitness can be accounted for by introducing the fitnesses $w_A=1+\tilde s$ and $w_B=1$ for alleles $A$ and $B$, respectively. 
In this case, the probability $p_{\rm r}(x)$ that a new individual carries allele $A$ after reproduction is
\be\label{prWF}
 p_{\rm r}(x)=\frac{w_A\Omega_A}{w_A\Omega_A+w_B\Omega_B}=\frac{(1+\tilde s)x}{1+\tilde s x}.
\ee

Now consider a structured population of $N$ (sub)populations (demes) of equal size $\Omega$ which form a fully-connected graph and in which inter-deme \emph{migration} can occur: the $i$-th deme is characterized by an allele frequency $x_i$ and, at each time step, it exchanges $m'=m\Omega$ individuals with every other deme of the population. 
Equivalently, it exchanges  $m'N$ individuals with a fictitious population whose allele frequency is $\bar x=\sum_{j=1}^{N} x_j/N$.
In terms of the frequencies $x_i$ and $\bar{x}$, the probability $p_{\rm m}(x_i,\bar x)$ that after migration an individual of the $i$-th deme carries allele $A$ is 
\be\label{pmWF}
 p_{\rm m}(x_i,\bar x)=m \bar x+(1-m)x_i,
\ee
in which the first term is the contribution of individuals coming from the other demes, while the second one accounts for those which did not move from 
the $i$-th deme. 

In the Wright-Fisher model, migration precedes reproduction and they take place in two subsequent steps 
\be
 (x_i,\bar x)\ \overset{\rm migr}{\longrightarrow}\ x_i'\ \overset{\rm repr}{\longrightarrow}\ x_i'',
 \label{sch:processes}
\ee
after which the initial frequency $x_i$ is changed into $x_i''$ and the  probability that an individual of the  $i$-th deme carries the allele $A$ is
\be
\label{pWF}
p(x_i,\bar x)=p_{\rm r}\big(p_{\rm m}(x_i,\bar x)\big).
\ee
Note that by inverting the order of these two pocesses this probability would be $p_{\rm m}(p_{\rm r}(x_i),\bar x)$, which is equivalent to $p(x_i,\bar x)$ in 
Eq.~\eqref{pWF} only for $m \ll 1$. In fact, in the reversed order, selection acts only on a fraction $1-m$ of the population and therefore the population always behaves like a neutral one in the limit $m\to 1$, see Eq.~\eqref{pmWF}. 
For this reason, hereafter we focus on the model in which migration precedes reproduction, as schematically indicated in~\eqref{sch:processes}.

For sufficiently large values of $\Omega$, one can rely on the diffusion approximation which considers only the mean variation $ \langle \Delta x_i\rangle$  and the mean square variation $ \langle \Delta x_i^2\rangle $ of the single-deme allele frequency which result from the implementation of the dynamical steps discussed above. For a binomial sampling with the probability $p(x_i,\bar x)$ given in Eq.~\eqref{pWF}, one finds
\be
\begin{cases}
\mu(x_i)  \equiv  \langle \Delta x_i\rangle =\langle x_i''-x_i\rangle=p(x_i,\bar x)-x_i=\tilde s x_i(1-x_i)+m(\bar x-x_i)+O(\tilde s^2,\tilde s m),\\[2mm]
v(x_i)  \equiv  \langle \Delta x_i^2\rangle =\langle (x_i''-x_i)^2\rangle= [x_i(1-x_i)+O(m,\tilde s)]/\Omega+O(\tilde s^2,m^2,m\tilde s).
\end{cases}
\label{eq:mu-v-WF}
\ee
Accordingly, within the diffusion approximation and for sufficiently small rates $\tilde s$ and $m$,  the dynamics of the Wright-Fisher model is described by the single-deme Langevin equation (1) presented in the Letter (which has to be interpreted with the It\^o prescription).

\subsection{Moran model}

In addition to the Wright-Fisher model discussed above, the Moran model \cite{sMoran} is also commonly used in order to describe the evolution of a haploid population of $\Omega$ individuals, each of which carries either allele $A$ or $B$. Although the Moran and the Wright-Fisher models are implemented with different rules at the microscopic level, we show here that they are actually described by the same Langevin equation, at least within a suitable range of parameters.

In the absence of selection (neutral model), at each time step of the dynamics of the Moran model two individuals (not necessarily distinct) are chosen at random: one is selected for death and the other for reproduction. The former is then removed from the population and it is replaced by an exact copy of the latter.
Since individuals are randomly chosen, the probability $d_A$ with which individuals carrying allele $A$ are removed from the population and the probability $r_A$ with which they reproduce are given by $d_A=r_A= x=\Omega_A/\Omega$, while the analogous probabilities for the individuals carrying allele $B$ are  $d_B=r_B=1-x$.

Within the Moran model, a selective advantage (e.g., for allele $A$) can be accounted for by modifying the fitness functions $w_{A,B}$ of the alleles, for instance by setting $w_ A=1+\tilde s$ and $w_B=1$ such that the probability for an individual carrying allele $A$ to be chosen for reproduction becomes $r_A(x)=(1+\tilde s)x/(1+\tilde s x)$. With this probability, at each step of the dynamics the number of individuals carrying allele $A$ increases/decreases by one with rates
\be
\label{rateMoran1}
  W_{+1}=r_Ad_B=(1+\tilde s)x(1-x)/(1+\tilde s x)\quad \mbox{and} \quad
  W_{-1}=r_Bd_A=(1-x)x/(1+\tilde s x),
\ee
respectively.
As in the case of the Wright-Fisher model, migration can be introduced at each step of the dynamics of the present %Moran 
model by selecting and exchanging two individuals belonging to different demes with probability $m/N$. The rates in Eq.~\eqref{rateMoran1} are consequently affected as
 \be
 \begin{cases}
 \label{rateMoran2}
  W_{+1}= (1+\tilde s)x_i(1-x_i)/(1+\tilde s x_i)+m\bar x(1-x_i),\\[1mm]
  W_{-1}= (1-x_i)x_i/(1+\tilde s x_i)+m(1-\bar x)x_i.
 \end{cases}
\ee
The time evolution of the probability distribution $P(\{x_i\},t)$ of the deme frequencies $\{x_i\}_i$ can be determined from the corresponding master equation with the rates given by Eq.~\eqref{rateMoran2}. For large $\Omega$ and in the limit of continuous time $\delta t \to 0$ (where $\delta t$ denotes the time interval separating two consecutive steps), standard expansions, such as the Kramers-Moyal expansion \cite{sGardiner}, lead to the Fokker-Planck equation
\be\label{FP_Moran}
 \partial_t P(\{x_i\},t)=-\sum_{j=1}^N\partial_{x_j}[\mu(x_j)P(\{x_i\},t)]+\frac{1}{2}\sum_{j=1}^N\partial_{x_j}^2[v(x_j)P(\{x_i\},t)],
\ee
in which the drift $\mu$ and the variance $v$ are given by
\be\label{drift_var}
\begin{cases}
  \mu(x_i)=(W_{+1}-W_{-1})/(\Omega\,\delta t)=\frac{\tilde s}{2} x_i(1-x_i)+\frac{m}{2}(\bar x-x_i)+O(\tilde s^2),\\[2mm]
  v(x_i)=(W_{+1}+W_{-1})/(\Omega^2\,\delta t)=[x_i(1-x_i)+O(\tilde s,m)]/\Omega,
\end{cases}
\ee
where we have chosen the temporal step to be $\delta t = 2/\Omega$. 
With this choice of time scales, the resulting genetic drift $v(x_i)$ for small  $\tilde s$ and $m$ is the same as the one of the  Wright-Fisher model for a population of the same size, see Eq.~\eqref{eq:mu-v-WF}.
Note that, in order to find the same expression also for the drift $\mu(x_i)$, it is necessary to rescale the migration and the selection coefficients as $m\to 2m$ and $\tilde s\to2\tilde s$, respectively. 
Equation~\eqref{FP_Moran} is nothing but the Fokker-Planck equation associated with the set of $N$ single-deme Langevin equatios (1) considered in the Letter, which, as we argued above, also describe the dynamics of the Wright-Fisher model.

\section{Langevin equation for the inter-deme mean frequency $\bar x$}

The single-deme Langevin equation (Eq.~(1) in the Letter) can be used in order to determine the infinitesimal increment of the inter-deme mean frequency (IDMF) $\bar x$ as
\be
 \rmd\bar x=\sum_{i=1}^N\frac{\partial \bar x}{\partial x_i}{\rmd x_i}=\frac{s}{N}\sum_{i=1}^Nx_i(1-x_i)(x_*-x_i)\rmd t+\frac{1}{N}\sum_{i=1}^N\sqrt{\frac{x_i(1-x_i)}{\Omega}}\rmd w_i,
\ee
where $\rmd w_i$ indicate the increments of the independent Wiener processes driving the dynamics of each single deme.  Since the individual stochastic increments $[x_i(1-x_i)/\Omega]^{1/2} \rmd w_i/N$ are independent and Gaussian random variables  with variance $x_i(1-x_i)/(\Omega N^2) \rmd t$, their sum is a Gaussian random variable with variance $ \sum_{i=1}^N x_i(1-x_i) \rmd t/(\Omega N^2) = (\avg{x}-\avg{x^2})/(\Omega N) \rmd t$, where the overbar indicates the mean over the demes.
The Langevin equation (2) of the Letter follows immediately.
As discussed in the Letter, this Langevin equation for $\avg{x}$ involves higher-order moments $\avg{x^k}$ which can be approximated by functions of $\avg{x}$ if one assumes that there is a separation between the local time scale which characterizes the response of $x_i$ to a change in $\bar x$, and the global time scale of $\bar x$ which, depending on the values of $s'N$ and $m'$, is either governed by the deterministic or by the stochastic contribution to the evolution of $\bar x$.
In fact, under this assumption, $x_i$ is expected to quickly relax into a quasi-stationary distribution $P_{\rm qs}(x_i|\avg{x}(t))$ corresponding to the slowly-varying $\avg{x}(t)$, which changes because of migration. Accordingly, one can write down the following effective Langevin equation for $\bar x$,
\be\label{Langevin_general}
 \dot{\avg{x}}=M(\avg{x})+\sqrt{V(\avg{x})}\,\eta,
\ee
where $M(\bar x)$ and $V(\bar x)$ are given by Eq.~(3) in the Letter and are calculated on the basis of the specific form of $P_{\rm qs}(x_i|\bar x)$ which follows from solving Eq.~(1) of the Letter with a fixed $\avg{x}$. 

\subsection{Time scales associated with Eq.~\reff{Langevin_general}}

There are two time scales emerging from Eq.~\reff{Langevin_general}: a \emph{relaxation time} associated with the deterministic term $M(\bar x)$ and a \emph{fluctuation time} associated with the stochastic term controlled by $V(\bar x)$.

\paragraph{Relaxation time. ---} By neglecting the stochastic fluctuations in Eq.~\reff{Langevin_general} one obtains $\dot{ \bar x}=M(\bar x)=s_{\rm e} \bar x(1-\bar x)(x_*-\bar x)$, where we used the expression for $M(\bar x)$ from Eq.~(3) in the Letter, which is valid under the assumption that the separation of time scales discussed above (and in the Letter) holds. This deterministic drift can be expanded around the metastable value $\bar x= \hat x \simeq x_*$ and the linear contribution is responsible for  a relaxation towards the value $\bar x=x_*$ which occurs exponentially in time, with a time scale
\be\label{Trel}
T_{\rm rel}=\frac{1}{s_{\rm e}x_*(1-x_*)}.
\ee

\paragraph{Fluctuation time. ---} Equation~\reff{Langevin_general} can be rewritten as
\be\label{Langevin_xbarra_breve}
 \rmd\bar x=M(\bar x) \rmd t+\sqrt{V(\bar x)} \rmd w,
\ee
where $\rmd w$ is a Wiener process with unit variance. In order to associate a time scale $ T_{\rm fluct}$ to the diffusion-like contribution of fluctuations, we note that the variance $V[\bar x]\rmd t +O(\rmd t^2)$ of $\avg{x}$ resulting from Eq.~\reff{Langevin_xbarra_breve} becomes of order unity for 
\be\label{Tfluct}
 T_{\rm fluct}\simeq 1/V[\bar x]\simeq \frac{N_{\rm e}}{x_*(1-x_*)},
\ee
where we used again Eq.~(3) of the Letter (which is valid if there is a separation of time scales) and we took the optimal value $x_*$ as the typical value for the IDMF $\bar x$.

By using the expression for $N_{\rm e}$ and $s_{\rm e}$ reported in the Letter, the condition $T_{\rm fluct}\gg T_{\rm rel}$ becomes (cp. Eqs.~\reff{Tfluct} and \reff{Trel})
\be
s'N\gg 1+\frac{1}{m'},
\ee
which turns into $s'm'N = sm\Omega^2 N \gg 1$ for $m' \ll 1$. This condition must be satisfied for the existence of a collective metastable state in the large-$N$ limit. 

\paragraph{Separation of time scales. ---}
The separation of time scales assumed in the Letter --- which allows one to determine the quasi-stationary distribution of each single deme and then use it in order to calculate approximate expressions for the higher-order moments $\avg{x^k}$ --- amounts at requiring  that the time scale $T_{\rm migr}\simeq 1/m$, associated with the response of $x_i$ to a change in $\bar x$ is much shorter than the one which characterizes the dynamics of $\bar x$. Since the latter involves essentially two different time scales, i.e., $T_{\rm fluct}$ and $T_{\rm rel}$ discussed above, $T_{\rm migr}$ must be much shorter than both of them: 
\be\label{separation}
 T_{\rm migr}\ll {\rm min}\{T_{\rm fluct},T_{\rm rel}\}.
\ee
Under the assumption that this inequality holds --- which can be verified a posteriori --- $T_{\rm rel}$ and $T_{\rm fluct}$ are given by Eq.~\reff{Trel} and \reff{Tfluct}, respectively. Accordingly, the minimum on the r.h.s.~of the previous equation is $T_{\rm rel}$ for $Ns'>1+1/m'$ and $T_{\rm fluct}$ otherwise. In the former instance, Eq.~\reff{separation} becomes $T_{\rm migr}\ll T_{\rm rel}$, i.e., $s_{\rm e}/m\ll 1$ (where one can neglect the factor $x_*(1-x_*)$, which is $\simeq 1/4$ within the range of parameters considered in the Letter).
The remaining case $Ns'<1+1/m'$ amounts at requiring $Nm'[1+1/(2m')]\gg 1$, which is satisfied whenever $N\gg 1$. 

In summary, the separation of time scales discussed here requires $s_{\rm e}/m\ll 1$ for $Ns'>1+1/m'$, while it always holds
(when $N$ is large) for $Ns'<1+1/m'$. 

\subsection{Metastable state}\label{Metastable_section}

Equation~(2) of the Letter always admits $\bar x=0$ and $\bar x=1$ as stationary solutions (\emph{absorbing states}), because $\bar x\in \{0,1\}$ implies $x_i=\avg{x}$ for all demes and therefore  $\avg{x^k}=\avg{x}^k$, $k\geq 2$.
For  $N=\infty$ (i.e., in the absence of the noise), another stationary solution $\bar x=\hat x$ is possible, which however becomes metastable for finite $N$ and corresponds to the non-trivial solution of 
\be\label{stationary_cond}
\avg{\mu(x_i)}=0.
\ee
If one neglects deme-to-deme fluctuations, such that $\avg{x^k} = \avg{x}^k$, the mean drift $\avg{\mu(x_i)}$ is given by $\mu(\avg{x})$ and therefore the non-trivial solution $\hat x$ of Eq.~\reff{stationary_cond} is $\hat x=x_*$.
As explained in the Letter, for large $N$ the deme average $\avg{x^k}$ can be approximated with the average $\langle x_i^k\rangle_{\rm qs}$ over a quasi-stationary distribution $P_{\rm qs}(x_i|\bar x)$ conditioned to $\bar x$. Equation~\reff{stationary_cond} then becomes
\be\label{stationary_cond2}
 \int_0^1{\rm d}x_i\, x_i(1-x_i)(x_*-x_i)P_{\rm qs}(x_i|\hat x)=0.
\ee
Substituting into Eq.~\reff{stationary_cond2} 
the stationary solution of the Fokker-Planck equation associated with Eq.~(1) in the Letter, i.e., $P_{\rm qs}(x_i|y)\propto x_i^{2m'y-1}(1-x_i)^{2m'(1-y)-1}\exp[s'x_i(2x_*-x_i)]$ (where $y$ is a function of $\bar x$ to be determined self-consistently from the condition $\langle x_i\rangle_{\rm qs}=\bar x$),
we find
\be\label{hat_x}
\hat x= x_*+\frac{1}{m'}\left(x_*-\frac{1}{2}\right)+O(s',1/m'^2),
\ee
for small selection rate $s'$ and large migration rate $m'$.

Note that $\hat x$ coincides with $x_*$ for $x_*=1/2$, as one can infer  from a detailed analysis of Eq.~\reff{stationary_cond2}, beyond the perturbative expansion in Eq.~\reff{hat_x}.
Upon moving $x_*$ away from $1/2$ towards one of the two boundary values 0 or 1, $\hat x$ moves in the same direction but with larger deviations with respect to $1/2$, such that it  reaches the boundary before $x_*$ does. 
This effect becomes more pronounced as the migration rate decreases, while, as shown by Eq.~\reff{hat_x},  $\hat x\simeq  x_*$  for large $m'$.
When $\hat x \in \{0,1\}$, the metastable state identified and discussed in the Letter does no longer exist and the dynamics of the whole population proceeds quickly to fixation. Interestingly enough, this occurs because the solution of the deterministic equation obtained from Eq.~\reff{Langevin_general}  in the limit $N\to\infty$, i.e., with $V = 0$ and therefore no noise, rapidly approaches the boundary values within a time which is independent of $N$ and even a small fluctuation is sufficient to cause fixation. On the contrary, when $\hat x \notin \{0,1\}$, the  state $\hat x$ is actually stationary for $N\to \infty$ and therefore its "lifetime" is expected to diverge as $N$ increases. This, in fact, qualifies the latter state as being metastable.

\section{Corrections to Eq.~(3) of the Letter}

The solution of the stationary Fokker-Planck equation associated with Eq.~(1) of the Letter under the assumption of a constant $\avg{x}$ and for vanishing selection $s=0$ is known to be the Beta distribution \cite{scherry-wakeley}
\be\label{Peq}
 P_{\rm qs}(x_i|y)=\frac{x_i^{2m'y-1}(1-x_i)^{2m'(1-y)}}{B(2m'y,2m'(1-y))},
\ee
where $B(u,v)$ in the normalization is the Beta function which can be expressed in terms of Euler's gamma function $\Gamma(u)$ as $B(u,v) = \Gamma(u)\Gamma(v)/\Gamma(u+v)$.
In the presence of selection,  an additional factor $\rme^{s'x_i(2x_*-x_i)}$ appears on the r.h.s.~of Eq.~\reff{Peq} and the associated normalization constant changes accordingly.  
The resulting distribution, including the first-order correction in $s$ (more precisely in $\alpha=s_{\rm e}/m$), for $x_*=1/2$ is given by
 \be\label{Pqs1}
  P_{\rm qs}^{(1)}(x|y)=\frac{x^{2m'y-1}(1-x)^{2m'(1-y)-1}}{B(2m'y,2m'(1-y))}\left\{ 1+\alpha(m'+1)\left[ \frac{2m'+1}{2m'}x(1-x)-y(1-y) \right] \right\}
+O(\alpha^2),
 \ee
where the correction term results from the expansion at the first order in $\alpha$ of the exponential in both the numerator and the normalization constant.
Equation~\reff{Pqs1} allows one to evaluate the moments $\langle x^k\rangle_{\rm qs} = \int_0^1\!\!\rmd x_i\, x_i^k   P_{\rm qs}^{(1)}(x_i|y)$ which --- within the assumptions on the existence of the metastable state discussed above --- can be used in order to calculate $\avg{x^k}$ for large $N$.
After the substitution of the parameter $y$ with the value $y(\avg{x})$ obtained as described in the Letter from the self-consistency condition $\avg{x} = \langle x\rangle_{\rm qs}$, we can use the moments $\langle x^k\rangle_{\rm qs}$ as an estimate of $\avg{x^k}$ which appears in the original Langevin equation for $\avg{x}$, reported as Eq.~(2)  in the Letter. As a result, this equation becomes
 \be\label{Langevin_xbarra_first_order}
 \begin{split}
 &\dot{\bar x}=s_e\bar x(1-\bar x)\left(\frac{1}{2}-\bar x\right)\Big\{ 1+\alpha[A+B\bar x(1-\bar x)]+O(\alpha^2) \Big\} \\
 &\hspace{3cm} +
 \sqrt{\frac{\bar x(1-\bar x)\left\{1+\alpha[C+D\bar x(1-\bar x)]+O(\alpha^2)\right\}}{N_e} }\, \eta,
 \end{split}
 \ee
  with
 \be
 \label{eq:corrections}
   A=\frac{1-7m'-6m'^2}{4m'(m'+2)(2m'+3)}, \quad 
   B=\frac{3(4+3m')}{(m'+2)(2m'+3)}, \quad 
   C=\frac{1-2m'}{4m'(2m'+3)}, \quad 
   D=\frac{3}{2m'+3}.
  \ee
As anticipated in the Letter, Eq.~\reff{Langevin_xbarra_first_order} takes the form of Eq.~\reff{Langevin_general}, where $M$ and $\sqrt{V}$ can be read by comparing the latter with Eq.~\reff{Langevin_xbarra_first_order} and render those reported in Eq.~(3) of the Letter for $\alpha=0$.
In Fig.~\ref{fig:M} we report the corresponding functions $M(\avg{x})/s_{{\rm e}}$ (panel (a)) and $N_{{\rm e}} V(\avg{x})$ (panel (b)) as functions of $\bar x$ for $m'=1$ and various values of $\alpha$. 
By comparing with the case $\alpha=0$ (solid line) one clearly sees that the first-order correction in $\alpha$ does not introduce new qualitative features in $M$ and $V$ but is merely responsible for some quantitative changes.
%
%%%
 \begin{figure}[h!]
  \flushleft
  \includegraphics[scale=0.47]{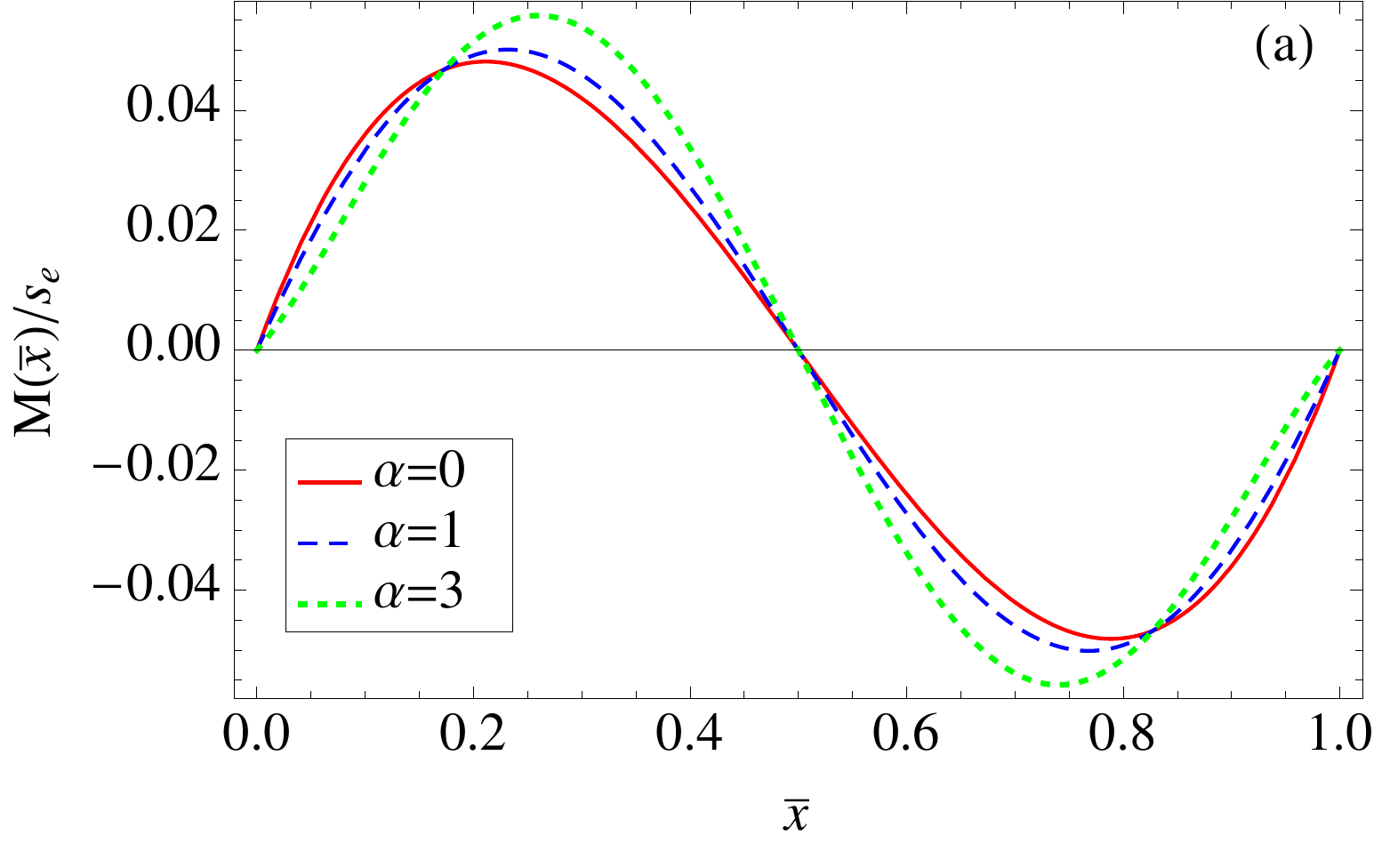}
  \includegraphics[scale=0.47]{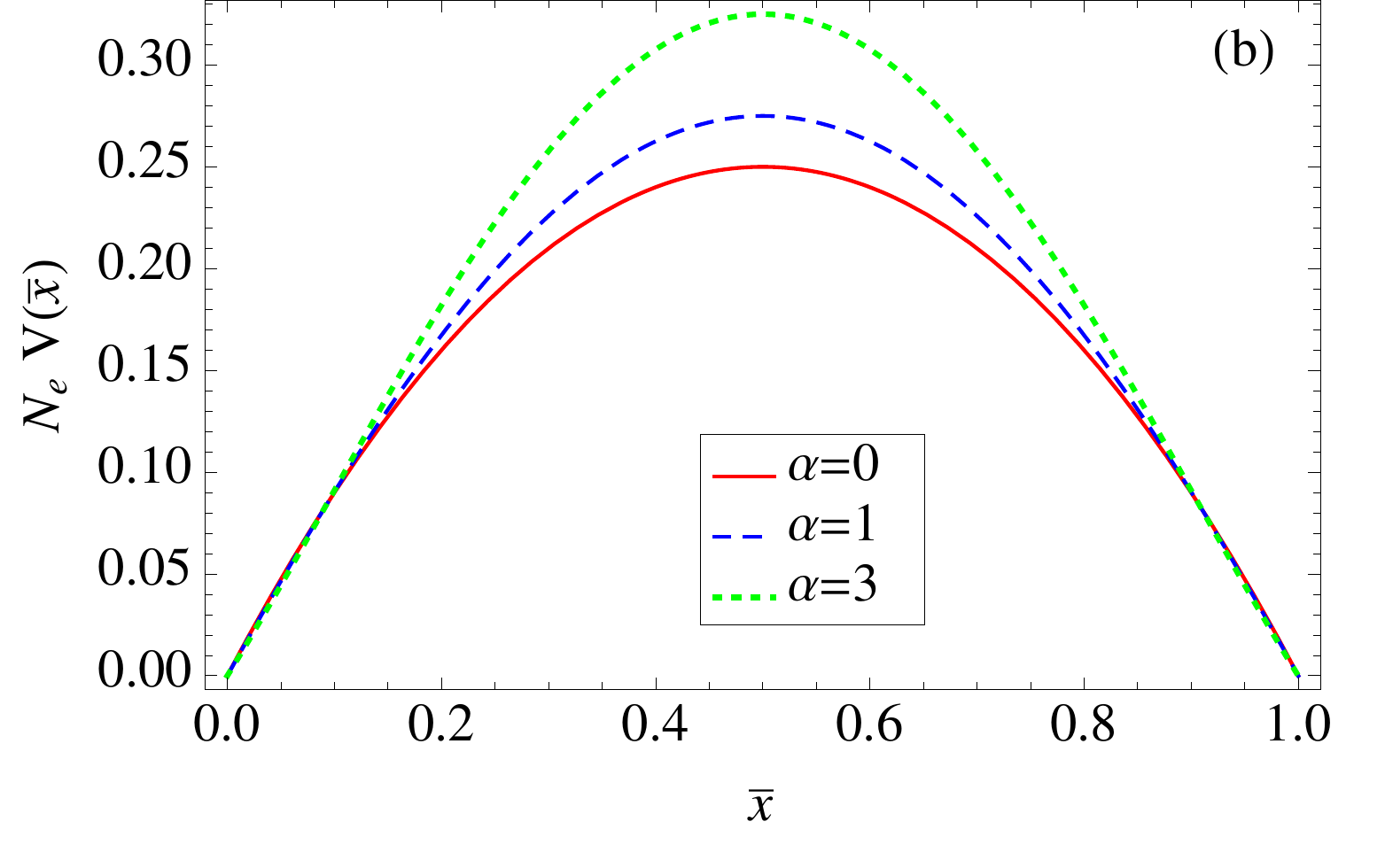}
  \caption{(a) Mean variation $M(\avg{x})$ and (b) variance $V(\avg{x})$ of the Langevin equation for the IDMF $\bar x$, as functions of $\bar x$, including the first-order correction in  $\alpha = s_{{\rm e}}/m'$ (see Eqs.~\reff{Langevin_xbarra_first_order} and \reff{eq:corrections}), for $m'=1$ and for various values of $\alpha$.
  }\label{fig:M}
 \end{figure}
%%%
%  

\subsection{Correction to the mean fixation time}

As a consequence of the corrections $O(\alpha)$ to the Langevin equation for $\avg{x}$, the corresponding mean fixation time (MFT) is modified compared to the value $T^{(0)}_{\rm fix}$ it has in the absence of these corrections (see Eq.~(5) of the Letter). 
In order to write the MFT in a compact form, it is convenient to perform first the change of variable $z=4x(1-x)$ in Eq.~\reff{Langevin_xbarra_first_order}: using the It\^o-Lemma and neglecting terms $O(\alpha^2)$, we obtain 
\be\label{Langevin_z_first_order}
\dot z =\widetilde{A}\,z(1-z)(1+\widetilde{C}z)-\widetilde{B}z(1+\widetilde{D}z)+\sqrt{4\widetilde{B}\,z(1-z)(1+\widetilde{D}z)}\,\eta,
\ee
with
\be
   \widetilde A=s_e(1+\alpha A)/2,\quad
   \widetilde B=(1+\alpha C)/N_e,\quad 
   \widetilde C= 4\alpha B/(1+\alpha A),\quad
   \widetilde D= \alpha D/(1+\alpha C).
 \ee
The MFT associated with Eq.~\reff{Langevin_z_first_order} can now be calculated via standard methods~\cite{sKimura} and its specific value depends on the initial condition of the system.
When the metastable state exists (i.e., for (i) $\bar x\simeq 1/2$ or $m'\gg 1$, necessary to have $\hat x\notin\{0,1\}$, and (ii) $Ns'\gg 1+1/m'$, necessary to have $\hat x$ metastable),
it is reached within a typical time $T_{\rm rel}$ which is largely independent of the size $N$ of the population and is much smaller than the MFT, which increases upon increasing the size $N$. Accordingly, for $N$ large enough, the specific choice $\bar x=x_0$ of the initial condition does not influence significantly the total elapsed time between the initial time of the dynamics and the fixation, provided that $x_0$ is far enough from the boundaries.
Assuming that the system starts from $\bar x = 1/2$ (corresponding to $z=1$), one finds
 \begin{equation}
  T_{\rm fix}^{(1)}=\frac{1}{2\widetilde B}\int_0^1\!\rmd z\,\frac{\rme^{-\beta z}}{\sqrt{1-z}\,(1+\widetilde Dz)^{\gamma}}\int_z^1\!\rmd\xi\frac{\rme^{\beta\xi}}{\xi\,\sqrt{1-\xi}\,(1+\widetilde D\xi)^{1-\gamma}},
 \end{equation}
where 
\be
\gamma=\widetilde A(\widetilde D-\widetilde C)/(2\widetilde B\widetilde D^2) \quad \mbox{and} \quad \beta=\widetilde A\widetilde C/(2\widetilde B\widetilde D).
\ee
The asymptotic behavior of $T_{\rm fix}^{(1)}$ for $m'\to 0$ can be easily calculated from the previous expression
\be\label{Tfix1}
\frac{T_{\rm fix}^{(1)}}{\Omega N}\simeq \frac{\log 2}{m'}\left(1-\frac{s'}{6}\right)
\ee
and it renders the one reported in the Letter for $s'=0$. For $s'\neq 0$, the negative correction on the r.h.s.~improves the agreement with the results of the simulations of the Wright-Fisher microscopic model (see Fig.~(2) of the Letter) compared to the theoretical prediction with $s'=0$.
The asymptotic expression 
of $T_{\rm fix}^{(1)}$ for large migration rate $m\to\infty$, instead,is the same as the one reported in the Letter at the lowest non-vanishing order in $s$:  
the population behaves like a well-mixed one with size $\Omega N$ and selection coefficient $s$.

\section{Global fixation in the absence of migration}\label{fixation_section}

In the absence of migration ($m=0$), each deme fixes independently of the others, but global fixation of the subdivided population occurs only when the last deme has fixed. The fixation probability $p(x_0,t)$ is defined as the probability to have $x(t)\in\{0,1\}$, assuming that the evolution of the stochastic variable $x$ started from $x(0)=x_0$ at time $t=0$; accordingly, $p$ is the cumulative distribution of the fixation times conditioned to the initial condition $x(0)=x_0$ \cite{sCrowKimura}.
Here we focus on the initial condition $\bar x_0=1/2$ (which approximately characterizes the metastable state), denoting by $P_N(t)$ the probability that all $N$ demes have already reached fixation at time $t$ with $x_i(0)=1/2$ for all of them.
Due to the independence of the demes (for $m=0$), this probability can be expressed in terms of the single-deme fixation probability $P_1(t)=p_{\rm fix1}(x_i=1/2,t)$ as $P_N(t)= P_1^N(t)$.
The probability density associated with the global fixation time is then given by $\dot P_N(t) = - \dot Q_N(t)$ where $Q_N(t) = 1 - P_N(t)$ and therefore, the average global fixation time is 
\be 
 T_{\rm fix}^{(m=0)}  =  \int_0^{\infty}\!\!\rmd t\,t\,\dot P_N(t) =  \int_0^{\infty}\!\!\rmd t\,Q_N(t) 
 =  \sum_{k=1}^N\binom{N}{k}(-1)^{k+1}\int_0^{\infty}
\!\!\rmd t\,Q_1^k(t),
\label{tfix-def} 
\ee
where $Q_1(t)= 1- P_1(t)$.
Assuming that the large fluctuations which cause fixation are independent Poisson processes, the probability that the system has not fixed after time $t$ is exponentially distributed around the average fixation time $T_{\rm fix1}$ of a deme, namely, $Q_1(t)\simeq \rme^{-t/T_{\rm fix1}}$. We checked numerically that this approximation is quite accurate in practice. 
Accordingly, from Eq.~\eqref{tfix-def} we find
\be
 T_{\rm fix}^{(m=0)}\simeq T_{\rm fix1}[\gamma+\psi(1+N)],
\ee
where $\gamma$ is the Euler constant and $\psi(z)$ is the digamma function, with an asymptotic behavior 
\be
 T_{\rm fix}^{(m=0)} \simeq T_{\rm fix1} \ln N \quad \mbox{for} \quad N \gg 1.
\ee
We point out that the MFT depends logarithmically on $N$, while in presence of migration such a dependence is at least linear (or even exponential, in the limit of large $N$).

\section{Numerical estimate of $\sigma_c$}

The lowest-order estimate $T_{\rm fix}^{(0)}$ of the MFT (Eq.~(5) of the Letter) can be written in terms of
\be
 \widetilde T_{\rm fix}^{(0)}(\sigma,m')\equiv\frac{T_{\rm fix}^{(0)}}{N\Omega}=\left(1+\frac{1}{2m'}\right)f(X),
\ee
where $X = \sigma m'/[4(m'+1)]$ and
\be\label{X}
f(X)=\int_0^1\rmd y \int_0^1 \rmd z\,\frac{\rme^{Xy(1-z^2)}}{1-yz^2}.
\ee
The stationary condition $\partial_{m'}T_{\rm fix}^{(0)} = 0 = \partial_{m'}\widetilde T_{\rm fix}^{(0)}(\sigma,m')$ therefore becomes 
an implicit equation in terms of $m'$ 
\be\label{minimo1}
 m'= 2[f(X)-Xf'(X)]/[2Xf'(X)-f(X)],
\ee
which admits  $m' = m'_{{\rm min}}(\sigma)$ as a solution.
Figure~3(b) of the Letter shows that, upon approaching the threshold value $\sigma_c$ of $\sigma$ at which $T_{\rm fix}^{(0)}(m')$ develops a non-monotonicity, $m'_{\rm min}(\sigma)$ diverges. By requiring  the r.h.s.~of Eq.~\reff{minimo1} to diverge for $\sigma\to\sigma_c$ we find numerically that $X_c\simeq 1.3$ and therefore $\sigma_c = 4 X_c \simeq 5.2$.

\section{Fixation probability}

The cumulative distribution of fixation times $p(\bar x,t)$, where $\bar x$ indicates the initial value of the IDMF, satisfies the backward Fokker-Planck (FP) equation 
\be\label{backFP}
\partial_t p(\bar x,t)= M(\bar x)\partial_{\bar x}p(\bar x,t)+V(\bar x)\partial_{\bar x}^2p(\bar x,t)/2,
\ee
where we assume the drift $M$  and variance $V$ given by Eq.~(3) of the Letter. We have solved a discretized version of Eq.~\reff{backFP}  on a grid in the $(\bar x,t)$ plane with spacings $(\Delta \bar x,\Delta t)$ given by $\Delta \bar x=1/500$ and $\Delta t$ ranging from $0.02$ to $0.001$, depending on the specific value of $m'$.
We checked numerically that the algorithm we employed for the solution of the differential equation converges upon decreasing suitably $\Delta \bar x$ and $\Delta t$.

Figure~\ref{fig:confronto} demonstrates that the probability  $p(\bar x=1/2,t)$  to reach fixation starting from the metastable state $\bar x=1/2$ evaluated from the numerical solution of Eq.~\reff{backFP} is, as a function of time $t$, quite well approximated  by an exponential law $\simeq 1 -\exp(-t/T_{\rm fix,fit})$,  for a suitable choice of $T_{\rm fix,fit}$. As a further test of the accuracy of the diffusive approximation also for determining rare events, we compare the solution of Eq.~\reff{backFP} with the results of numerical simulations of the  Wright-Fisher model described in Sec.~\ref{WFsection}. In particular, we computed the fixation time of the model by averaging over about 500 realizations of the dynamics. The resulting cumulative distribution is reported in Fig.~\ref{fig:confronto} (red solid line) and it displays a good agreement  with the numerical solution of Eq.~\reff{backFP} (blue dotted line).
The decay time $T_{\rm fix,fit}$ which characterizes the exponential law reported in Fig.~\ref{fig:confronto} (green dotted line) is chosen such that to fit the MFT resulting from the simulation of the Wright-Fisher model.  
As it can be seen from Fig.~\ref{B-fig:Tfix} in the Letter, within the range of parameters considered there, $T_{\rm fix,fit}$ computed from the WF simulations agrees rather well with $T_{\rm fix}^{(0)}$ determined according to Eq.~\reff{B-Tfix0} of the Letter.
%
%%%
\begin{figure}[h!]
  \centering
  \includegraphics[scale=0.63]{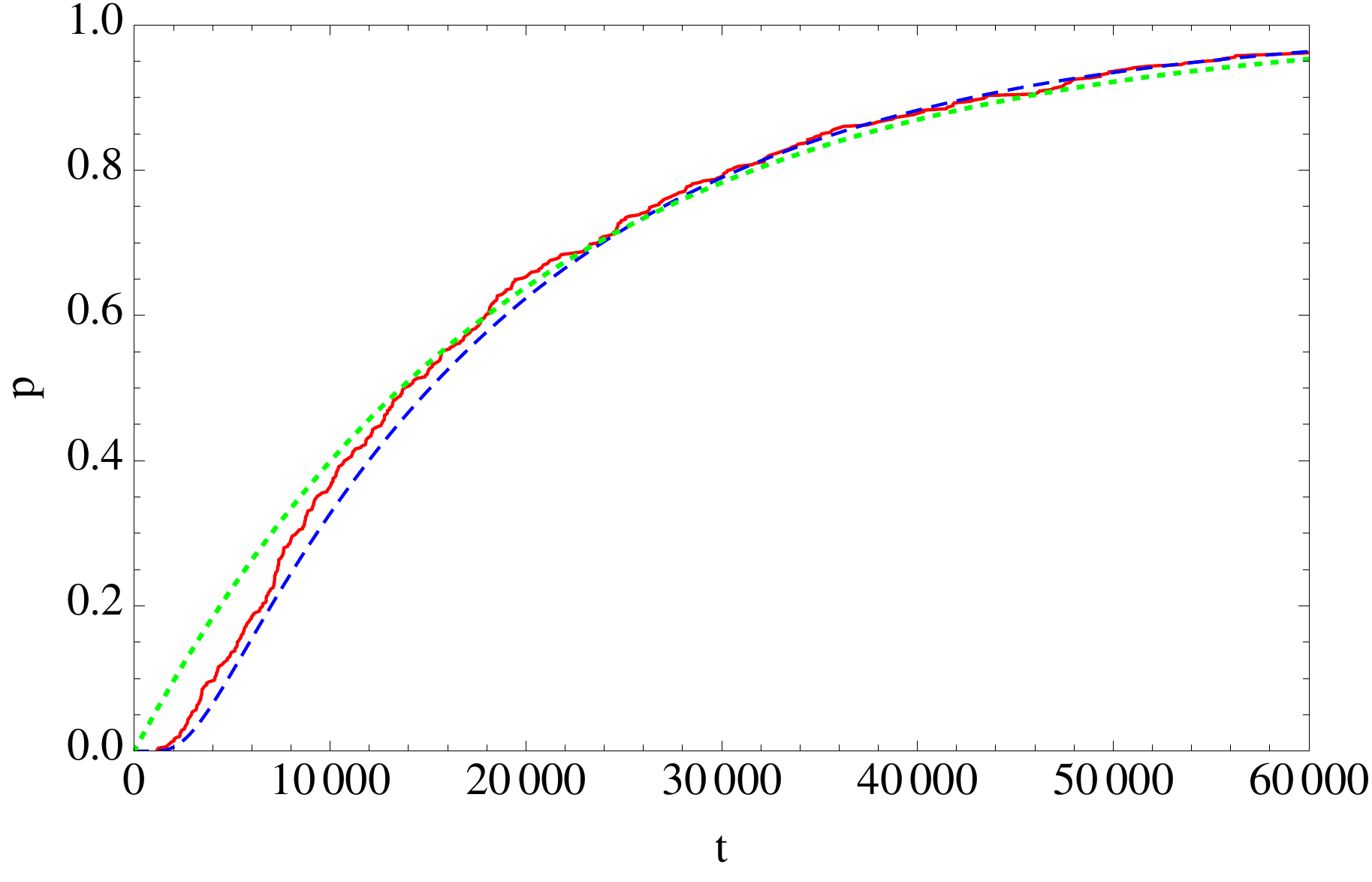}
  \caption{Fixation probability $p(\bar x,t)$ as a function of time, for a population which starts from the initial value $\avg{x}=1/2$. The numerical solution of Eq.~\reff{backFP} (blue, dashed line), is compared with the results of numerical simulations of the Wright-Fisher model (red, solid line) and with an exponential law $1 -\exp(-t/T_{\rm fix,fit})$ (green dotted). This plot refers to a population with $\Omega=100$, $N=30$, $s'=1$, $m'=0.3$, and $x_*=1/2$. 
  }\label{fig:confronto}
\end{figure}
%%%
%
We point out that, in the absence of migration, the single-deme fixation probability (i.e., the cumulative distribution of $T_{\rm fix1}$ discussed in Sec.~\ref{fixation_section}) satisfies Eq.~\reff{backFP}, where the functions $M(\avg{x})$ and $V(\avg{x})$ are given by Eq.~(3) of the Letter in which, however, the effective parameters $s_{\rm e}$ and $N_{\rm e}$ are replaced by $s$ and $\Omega$, respectively, which refer to the single deme. Accordingly, the resulting distribution of $T_{\rm fix1}$ has the same qualitative behavior as the fixation time discussed here, though with a different time scale.

\section{Intra-deme heterozygosity $h(t)$}

With a procedure analogous to the one described in the Letter for the global heterozygosity $H$, one can obtain an estimate for the time evolution of the  intra-deme heterozygosity $h$. The only difference compared to the case of $H$ is that the value of $h$ in the metastable state --- which is taken to be the initial condition in our heuristic estimate ---  is $h(0)=\hat h$, computed as follows.
We assume that at time $t=0$ each deme is distributed according to the quasi-stationary distribution $P_{\rm qs}(x_i | \hat x)$, where $\hat x$ is the value of $\avg{x}$ in the metastable state discussed in Sec.~\ref{Metastable_section}.
In the presence of balancing selection with $x_*=1/2$, one has $\hat x=1/2$ (independently of the values of $m'$ and $s'$) and the value $\hat h$  of the intra-deme heterozygosity $h$ in the metastable state is reported in Fig.~\ref{fig:hmet} as a function of the migration rate $m'$ for large $N$ (the actual behavior does not change much for smaller values of $N$).
%
%%%
 \begin{figure}[h!]
  \centering
  \includegraphics[scale=0.8]{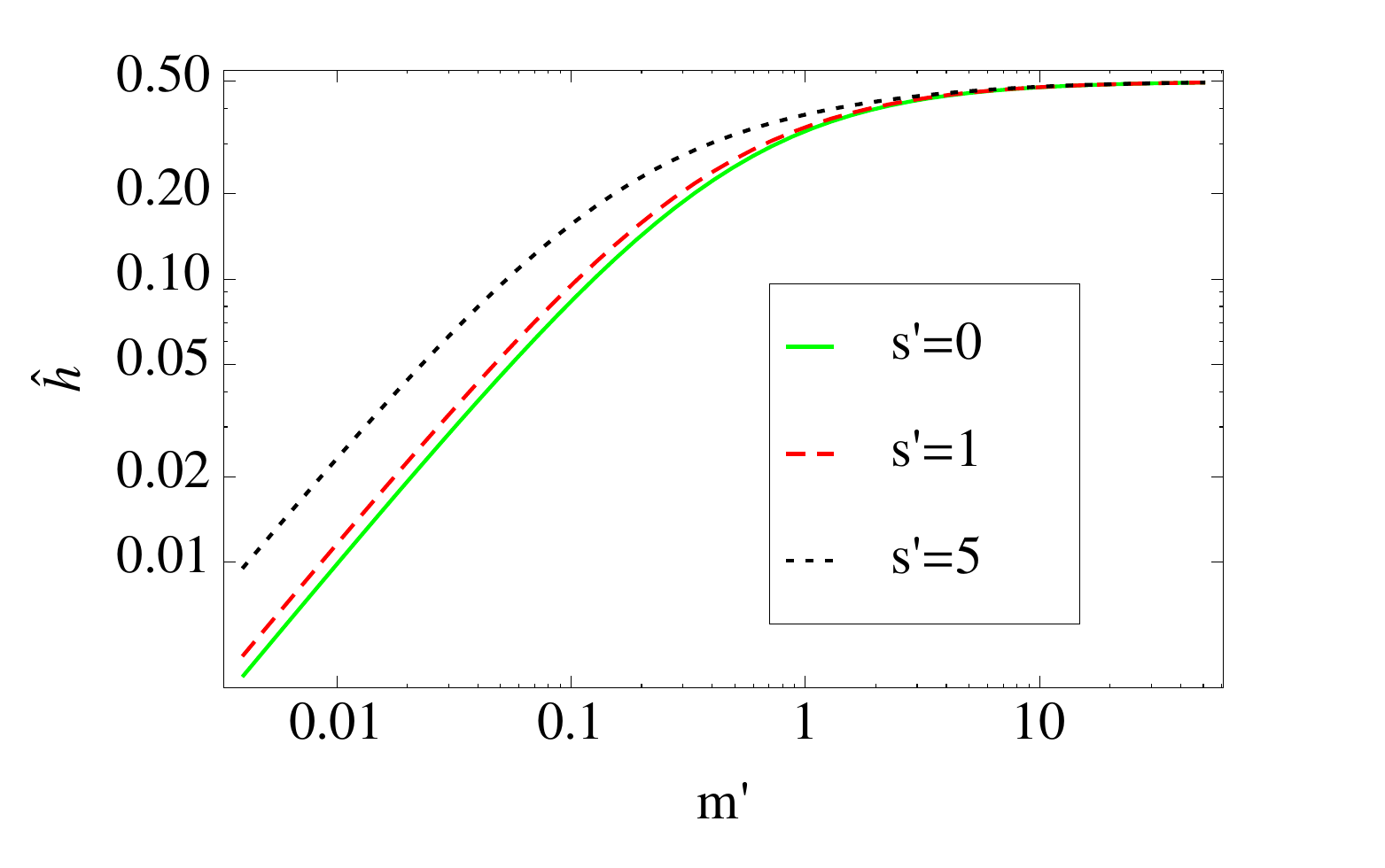}
  \caption{Intra-deme heterozygosity  $h$ within the metastable  state $x_i \simeq \bar x \simeq \hat x$ as a function of the (rescaled) migration rate $m'$ and for various values of the (rescaled) selection rate $s'$, in a population with $N=\infty$.}\label{fig:hmet}
 \end{figure}
%%%
%
On the basis of this initial value $\hat h$ of $h$, an estimate of $h(t)$ can be obtained under the same assumption as the one which was used in order to discuss $H(t)$. The corresponding evolution is reported in Fig.~\ref{fig:h(t)}. For slow and fast migration, $h(t)$ remains close to $h(0)$ for a long time,  whereas it rapidly decreases in time for intermediate values of the migration rate. For a sufficiently large time, instead, the profile of $h(t)$ as a function of $m'$ develops a minimum
at $m'\simeq m'_{\rm min}$. 
Our predictions agree rather well with the results of simulations, while they become less accurate for $m'\lesssim 1/\sigma\simeq 0.03$ which is outside the range of validity of our approximation.

%
%%%
 \begin{figure}[h!]
  \centering
  \includegraphics[scale=0.6]{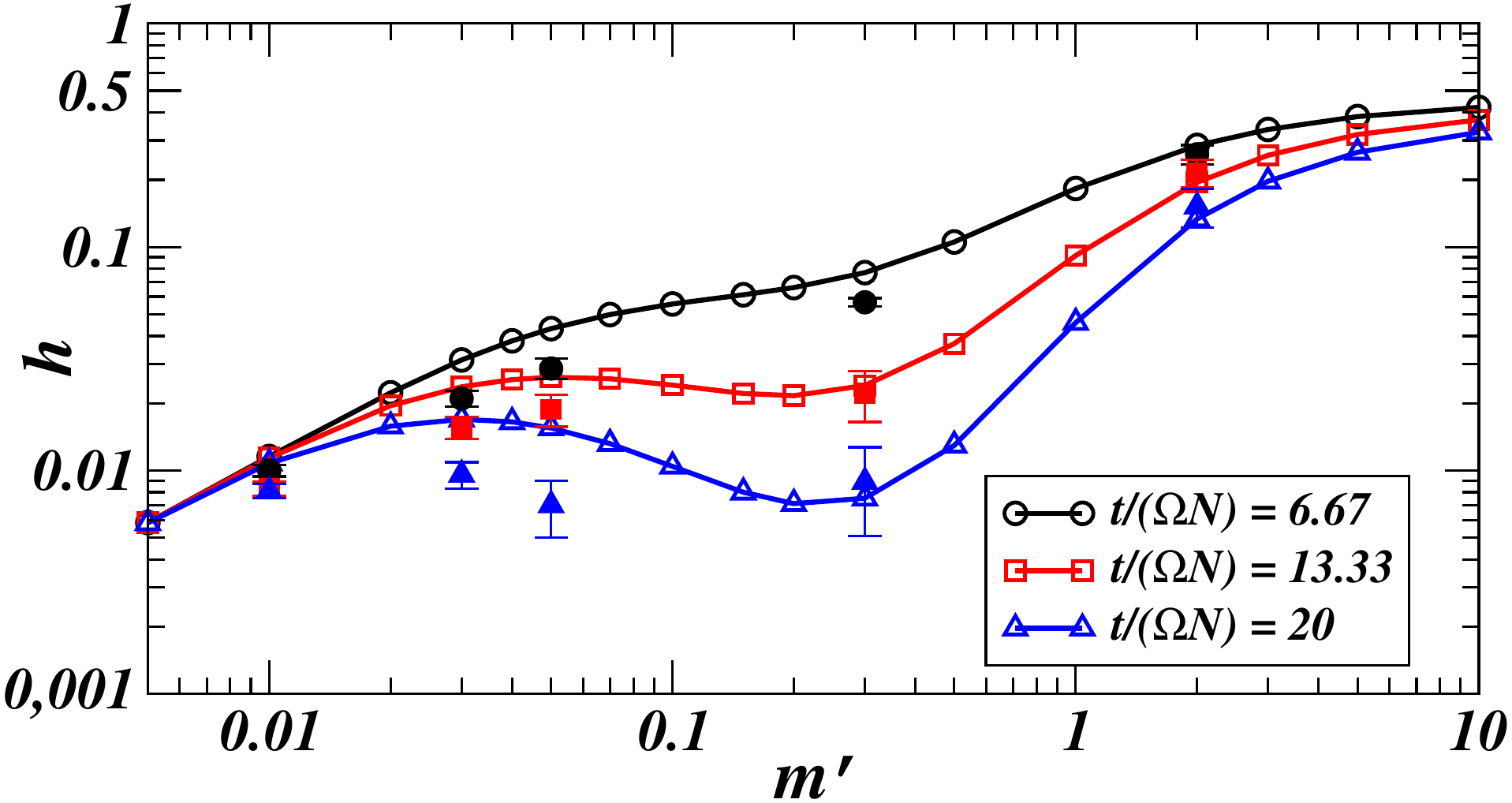}
  \caption{Intra-deme heterozygosity $h$  as a function of the migration rate $m'$ and at various times during the evolution of the subdivided population with $N=30$, $s'=1$, and $x_\ast = 0.5$. The evolution calculated on the basis of the approximation described in the main text (continuous lines) is compared with the results of the numerical simulations of the Wright-Fisher model (symbols with error-bars). The approximation is expected to become increasingly accurate as $T_{\rm rel}\ll T_{\rm fluct}$, i.e., as $m'\gg 1/\sigma \simeq0.033$.}\label{fig:h(t)}
 \end{figure}
%%%
%

\section{Bounds for the mean fixation time}

In this section we report the values of the bounds for the mean fixation time (MFT) $T_{\rm fix}$ which were derived in Ref.~\cite{sSlatkin} (Ref.~[11] of the Letter).

\subsection{Upper bound for slow migration}

In the limit of small migration rate $m$, the analysis of Ref.~\cite{sSlatkin} yields the following upper bound for the MFT
\be
 T_{\rm fix}(m\to0)\leq\frac{N}{\Omega\,m\,u_{1}(1/\Omega,\infty)},
     \label{eq:SM:Tfixsmall}
\ee
where $u_1(x_0,\infty)$ is the probability that, in the absence of migration, the generic deme  of the subdivided population (composed of $N$ identical demes) eventually reaches fixation in the absorbing state with $x=1$, starting from the initial condition $x=x_0$. 
Note that $u_1$ does not coincide with $p_{\rm fix1}$ defined in Sec.~\ref{fixation_section}, because the latter is the fixation probability to any of the two absorbing boundaries $x=0,1$. 
In fact, $u_1$ can be calculated with standard methods (see, e.g., Ref.~\cite{sKimuraGenetics}) which give
\be
  u_{1}(x,t\to\infty)=\frac{\int_0^x \rmd y\,\exp[-s'y(1-y)]}{\int_0^1\rmd y\,\exp[-s'y(1-y)]}
\ee
in the presence of balancing selection with $x_*=1/2$. 

\subsection{Limit of fast migration}

In the limit of large migration rate $m \to\infty$, the subdivided population is expected to behave as a well-mixed population with the same total number $\Omega N$ of individuals; accordingly  the MFT can be calculated by specializing the results of Ref.~\cite{sKimura} to the case of balancing selection with $x_*=1/2$:
\be
  T_{\rm fix}(m\to\infty)=\Omega N\int_0^1 \rmd y \int_0^1 \rmd z \frac{\rme^{s\Omega N y (1-z^2)/4}}{1- y z^2}.
  \label{eq:SM:Tfixlarge}
\ee
It can be noticed from Eq.~\reff{eq:SM:Tfixlarge} that the MFT for large migration rate strongly depends on $N$: if we fix all other parameters, $T_{\rm fix}(m\to\infty)$ increases exponentially as a function of $N$.
In addition, even though the expressions for the  bounds reported in Eqs.~\reff{eq:SM:Tfixsmall} and \reff{eq:SM:Tfixlarge} are specific to the case of balancing selection with $x_*=1/2$, they can be easily generalized as we did in Fig.~3(c) of the Letter, where the bound for $m\gg 1$ was reported also for $x_*\neq 0.5$.
Finally, we emphasize that our prediction for the MFT approaches the bound \reff{eq:SM:Tfixlarge} from below whenever it is a non-monotonic function of $m$ while it does so from above --- as expected from Ref.~\cite{sSlatkin} --- when such a non-monotonicity is absent.

\end{widetext}

\end{document}